\documentclass{article}
\usepackage{amssymb}
\usepackage{amsmath, amsfonts}

\begin{document}

\begin{center}
{\Large Some connections between binary block codes and Hilbert algebras}

\begin{equation*}
\end{equation*}%
Cristina FLAUT%
\begin{equation*}
\end{equation*}

{\small Faculty of Mathematics and Computer Science,}

{\small Ovidius University,}

{\small Bd. Mamaia 124, 900527, Constan\c{t}a,}

{\small Rom\^{a}nia}

{\small cflaut@univ-ovidius.ro}

{\small cristina\_flaut@yahoo.com}

{\small http://cristinaflaut.wikispaces.com/}

{\small http://www.univ-ovidius.ro/math}/%
\begin{equation*}
\end{equation*}
\end{center}

\textbf{Abstract.} {\small In this paper, we will study some connections
between Hilbert algebras and binary block-codes.With these codes, we can
eassy obtain orders which determine suplimentary properties on these
algebras. We will try to emphasize how, using binary block-codes, we can
provide examples of classes of Hilbert algebras with some properties, in our
case, classes of semisimple Hilbert algebras and classes of local Hilbert
algebras. }%
\begin{equation*}
\end{equation*}

\bigskip


\begin{equation*}
\end{equation*}

\textbf{1. Introduction}%
\begin{equation*}
\end{equation*}

Over the last years, codes have experienced a significant development. Using
codes, an impressive quantity of data can be transmitted. They~have
important applications in various domains with implications in social life.
Using codes, data are trasformed into a form which can be easily understand
by computer software and can be represented in a form which is more
resistant to errors in data transmission or data storage. In this way, data,
in their quantitative or qualitative forms, are classified to facilitate
some analysis.

Coding Theory is a mathematical domain with many applications in Information
Theory. Various type of codes and their connections with other mathematical
objects have been intensively studied. One of these applications, namely
connections\ between binary block codes and BCK-algebras, was \ recently
studied in (Jun,  Song, 2011), (Flaut, 2015) and (Borumand Saeid,
Fatemidokht, Flaut, Kuchaki Rafsanjani, 2015). Starting from these results
and since a positive implicative BCK-algebra is a Hilbert algebra, in this
paper we will study some connections between Hilbert algebras and binary
block-codes. Even if in (Borumand Saeid, Fatemidokht, Flaut, Kuchaki
Rafsanjani, 2015) and (Chajda, Hala\v{s}, Jun, 2002) it was remarked that a
BCK-algebra and a Hilbert algebra can be obtained on any ordered set with a
greatest element $\theta ,$ using codes we can easily obtain orders which
determine suplimentary properties of these algebras. In the following, we
will try to emphasize how, using binary block-codes, we can provide examples
of classes of Hilbert algebras with certain properties, in our case, classes
of semisimple Hilbert algebras (Theorem 3.11) and classes of local Hilbert
algebras (Theorem 3.12). Hilbert algebras were first introduced at the
middle of the 20th century and they are an important tool for some
investigations in intuitionistic logics and other non-classical logics (see
(Piciu, Bu\c{s}neag,  2010)). 
\begin{equation*}
\end{equation*}

\textbf{2. Preliminaries}%
\begin{equation*}
\end{equation*}
\textbf{Definition 2.1.} An algebra $(X,\ast ,\theta )$ of type $(2,0)$ is
called a \textit{BCI-algebra} if the following conditions are fulfilled:

$1)~((x\ast y)\ast (x\ast z))\ast (z\ast y)=\theta ,$ for all $x,y,z\in X;$

$2)~(x\ast (x\ast y))\ast y=\theta ,$ for all $x,y\in X;$

$3)~x\ast x=\theta ,$ for all $x\in X$;

$4)$ For all $x,y,z\in X$ such that $x\ast y=\theta ,y\ast x=\theta ,$ it
results $x=y$.

If a BCI-algebra $X$ satisfies the following identity:

$5)$ $\theta \ast x=\theta ,~$for all $x\in X,$ then $X$ is called a \textit{%
BCK-algebra}.

A BCK-algebra $X$ is called \textit{commutative }if $x\ast (x\ast y)=y\ast
(y\ast x),$ for all $x,y\in X$ and \textit{implicative }if $x\ast (y\ast
x)=x,$ for all $x,y\in X.$ A BCK-algebra $(A,\ast ,0)$ is called \textit{%
positive implicative} if and only if%
\begin{equation*}
\left( x\ast y\right) \ast z=\left( x\ast z\right) \ast (y\ast z),\text{ for
all }x,y,z\in A.
\end{equation*}

The partial order relation on a BCK-algebra is defined such that $x\leq y$
if and only if $x\ast y=\theta .$

\textbf{Remark 2.2.} The following BCK-algebra $(X,\ast ,\theta )$%
\begin{equation}
\begin{array}{c}
\theta \ast x=\theta \text{ and }x\ast x=\theta ,\forall x\in X; \\ 
x\ast y=\theta ,\text{ if \ }x\leq y,\ \ \ x,y\in X; \\ 
x\ast y=x,\text{ otherwise }%
\end{array}
\tag{2.1.}
\end{equation}%
is a non-commutative and a non-implicative algebra. (see (Diego,1966) and
(Flaut, 2015))\medskip 

\textbf{Definition 2.3.} \ A \textit{Hilbert algebra} is a triplet $(H,\ast
,1)$ in which $H$ is a non-empty set, "$\ast $" a binary operation on $H$
and $1\in H$ is a fixed element such that the following relations hold, for
all $x,y,z\in H$:

1) $x\ast \left( y\ast x\right) =1;$

2) $\left( x\ast \left( y\ast z\right) \right) \ast \left( \left( x\ast
y\right) \ast \left( x\ast z\right) \right) =1;$

3) $x\ast y=1$ and $y\ast x=1$ imply $x=y.\medskip $

\textbf{Proposition 2.4.} ((Dudek, 1999), Theorem) \textit{An algebra} $%
(H,\ast ,\theta )$ \textit{is a Hilbert algebra if and only if its dual
algebra} $\left( H,\cdot ,\theta \right) ,$\textit{where} $x\cdot y=y\ast x,$
\textit{is a positive implicative BCK-algebra.}\medskip 

\textbf{Definition 2.5. }Let $\left( H,\cdot ,\theta \right) $ be a Hilbert
algebra. A subset $L$ of the algebra $H$ is called a \textit{filter }(or
implicative filter) of $H$ if we have $\theta \in L$ and if for all $x,y\in
L,$ from $x\in L$ and $x\cdot y\in L$ it results that $y\in L.\medskip ~$

\textbf{Definition 2.6.} (see (Bu\c{s}neag, 1987)) Let $H$ be a Hilbert
algebra.

1) Let $L$ be a filter of the algebra $H.$ $L$ is called \textit{maximal} if
for a proper filter $F$ of $H$ if $L\subseteq F,$ we have $L=F.$

2) $H$ is called a \textit{semisimple} Hilbert algebra if the intersection
of all maximal filters of $H$ is $\{\theta \}$.

3) $H$ is called \textit{a local algebra} if and only if the algebra $H$
contains only one maximal filter.

For other details about Hilbert algebras, the reader is referred to ( Bu\c{s}%
neag, 1987), (Dan, 2008).%
\begin{equation*}
\end{equation*}

\textbf{3. Main results}%
\begin{equation*}
\end{equation*}

\textbf{Proposition 3.1.} \textit{BCK-algebra} $(X,\ast ,\theta )\ \ $%
\textit{defined by the relation} $\left( 2.1.\right) $\textit{\ is a
positive implicative algebra.\smallskip }

\textbf{Proof. }We must prove that $\left( x\ast y\right) \ast z=\left(
x\ast z\right) \ast (y\ast z),$ for all $x,y,z\in X.$

Case 1: at least one element is\ $\theta .$ \newline
i)$~\left( \theta \ast x\right) \ast z=\theta \ast z=\theta $ and $\left(
\theta \ast z\right) \ast (x\ast z)=\theta \ast (x\ast z)=\theta ;$\newline
ii) $\left( x\ast \theta \right) \ast z=x\ast z=x$ and $\left( x\ast
z\right) \ast (\theta \ast z)=x\ast \theta =x;$\newline
iii) $\left( x\ast y\right) \ast \theta =x\ast y=x$ and $\left( x\ast \theta
\right) \ast (y\ast \theta )=x\ast y=x;$\newline

Case 2: one element is comparable with another.\newline
i) $x\leq y;$ $\left( x\ast y\right) \ast z=\theta \ast z=\theta $ and $%
\left( x\ast z\right) \ast (y\ast z)=x\ast y=\theta ;$\newline
ii) $x\leq z;$ $\left( x\ast y\right) \ast z=x\ast z=\theta $ and $\left(
x\ast z\right) \ast (y\ast z)=\theta \ast y=\theta ;$\newline
iii) $z\leq y;$ $\left( x\ast y\right) \ast z=x\ast z=x$ and $\left( x\ast
z\right) \ast (y\ast z)=x\ast y=x;$\newline
iv) $y\leq x;$ $\left( x\ast y\right) \ast z=x\ast z=x$ and $\left( x\ast
z\right) \ast (y\ast z)=x\ast y=x;$\newline
v) $z\leq x;$ $\left( x\ast y\right) \ast z=x\ast z=x$ and $\left( x\ast
z\right) \ast (y\ast z)=x\ast y=x;$\newline
vi) $y\leq z;$ $\left( x\ast y\right) \ast z=x\ast z=x$ and $\left( x\ast
z\right) \ast (y\ast z)=x\ast \theta =x;$

Case 3: two elements are comparable with the third.\newline
$x\leq y$ and $z\leq y;\left( x\ast y\right) \ast z=\theta \ast z=\theta $
and $\left( x\ast z\right) \ast (y\ast z)=x\ast y=\theta ,etc.\Box \medskip $

\textbf{Remark 3.2.} Let $v_{x}=x_{1}x_{2}\ldots x_{n}$ and $%
v_{y}=y_{1}y_{2}\ldots y_{n}$ be two codewords belonging to a binary
block-code $V$. We define an order relation $\leqslant _{c}$ on the set of
codewords belonging to a binary block-code $V,$ as follows ( see (Jun, Song,
2011)):

\begin{equation}
v_{x}\leqslant _{c}v_{y}\Leftrightarrow y_{i}\leqslant x_{i}\ \text{for \ }%
i=1,2,\ldots ,n.\medskip \newline
\tag{3.1.}
\end{equation}

\textbf{Definition 3.3.} (Jun, Song, 2011)

i) A mapping $f:A\rightarrow X$ is called a \textit{BCK-function} on $A.$ A 
\textit{cut function} \textit{of} $f$ is a map $f_{r}:A\rightarrow
\{0,1\},r\in X,$ such that $f_{r}\left( x\right) =1,$ if and only if \ $%
r\ast f\left( x\right) =\theta ,\forall x\in A.$ \newline
A \textit{cut subset} of $A$ is the following subset of $A,$ $A_{r}=\{x\in
A:r\ast f\left( x\right) =\theta \}.$

ii) Let $A=\{1,2,\ldots ,n\}$ and let $X$ be a BCK-algebra. For each
BCK-function $\tilde{A}:A\rightarrow X$  a binary block-code of length $n$
was defined. A codeword in a binary block-code $V$ is $v_{x}=x_{1}x_{2}%
\ldots x_{n}$ such that $x_{i}=x_{j}\Leftrightarrow A_{x}(i)=j,$ for $i\in A$
and $j\in \{0,1\}$.\medskip \medskip 

Let $X$ be a BCK-algebra. Let $V$ be a  binary block-code with $n$ codewords
of length $n.$ We consider the matrix $M_{V}=\left( m_{i,j}\right) _{i,j\in
\{1,2,...,n\}}\in \mathcal{M}_{n}(\{0,1\})$ with the rows consisting of the
codewords of $V.$ This matrix is called \textit{the matrix associated to the
code} $V.\medskip $

Let $C$ be a binary block code with $n$ codewords of length $m.$ From
Proposition 3.8 and Theorem 3.9 from (Flaut, 2015), we find a BCK-algebra $X$
such that the obtained binary block-code $V_{X}$ contains the binary
block-code $C$ as a subset. In the following, we briefly present this
procedure.

Let $V$ be a binary block-code, $V=\{x_{1,}x_{2},...,x_{n}\},$ with
codewords of length $m.$ We consider the codewords $x_{1,}x_{2},...,x_{n}$
lexicographically ordered, $x_{1}\geq _{lex}x_{2}\geq _{lex}...\geq
_{lex}x_{n}.$ Let $M\in \mathcal{M}_{n,m}(\{0,1\})$ be the associated matrix
with the rows $w_{1},...,w_{n}$ in this order. We can extend the matrix $M$
to a square matrix $M^{\prime }\in \mathcal{M}_{p}(\{0,1\}),~p=n+m,$ such
that $M^{\prime }=\left( m_{i,j}^{\prime }\right) _{i,j\in \{1,2,...,p\}}$
is an upper triangular matrix with $m_{ii}=1,$ for all $i\in \{1,2,...,p\}.$
For this purpose, we insert in the left side of the matrix $M$ (from the
right to the left) the following $n$ new columns of the form $\underset{n}{%
\underbrace{00...01}},\underset{n}{\underbrace{00...10}},...,\underset{n}{%
\underbrace{10...00}}.$ A new matrix $D$ with $n$ rows and $n+m$ columns
results. Now, we insert at the bottom of the matrix $D$ the following $m$
rows: $\underset{n}{\underbrace{00...0}}\underset{m}{\underbrace{10...00}}$
, $\underset{n+1}{\underbrace{00...0}}\underset{m-1}{\underbrace{01...00}}%
,...,\underset{n+m-1}{\underbrace{000}}1.$ We obtain the required matrix $%
M^{\prime }.~$If the first line of \ the matrix $M^{\prime }$ is not $%
\underset{p}{\underbrace{11...1}},$ then we insert the row $\underset{p+1}{%
\underbrace{11...1}}$ as a first row and the column $1\underset{p}{%
\underbrace{0...0}}~$as a first column$.$ We obtain a new code $W$\ $%
=\{\theta ,w_{1},...,w_{n+m}\}.$ Using relation $\left( 2.1\right) ,$ we
define on $\left( W,\leqslant _{c}\right) $ a binary relation $"\ast "$. It
results that $X=\left( W,\ast ,\theta \right) $ becomes a
BCK-algebra.\medskip 

\textbf{Proposition 3.4.} \textit{Let} $C$ \textit{be a binary block code
with }$n$\textit{\ codewords of length} $m.$ \textit{With the above
notations, we have that }$\{\theta ,w_{n+1},...,w_{n+m}\}$\textit{\
determines a filter in the Hilbert algebra }$(X,\cdot ,\theta ),$ \textit{%
obtained in Remark 3.2.\smallskip }

\textbf{Proof.} It is obvious.$\medskip $

\textbf{Definition 3.5.} A mapping $\tilde{A}:A\rightarrow X$ is called an 
\textit{H-function on} $A$, where $A$ and $X$ are a nonempty set and $X$ is
a\ Hilbert-algebra, respectively.\medskip 

\textbf{Definition 3.6. }A cut function of $\tilde{A}$, for $q\in X$, \
where $X$ is a Hilbert algebra, is defined to be a mapping $\ \tilde{A}%
_{q}:A\rightarrow \{0,1\}$ such that $(\forall x\in A)(\tilde{A}%
_{q}(x)=1\Leftrightarrow q\ast \tilde{A}(x)=0).\medskip $

\textbf{Definition 3.7. }Let $A=\{1,2,\ldots ,n\}$ and let $X$ be a
Hilbert-algebra. A codeword in a binary block-code $V$ is $%
v_{x}=x_{1}x_{2}\ldots x_{n}$ such that $x_{i}=x_{j}\Leftrightarrow 
\widetilde{A}_{x}(i)=j$ for $i\in A$ and $j\in \{0,1\}$.\medskip

\textbf{Proposition 3.8.} \textit{Let} $C$ \textit{be a binary block code
with} $n$ \textit{codewords of length} $m$ \textit{and let} $X$ \ \textit{be
the associated Hilbert algebra, as in the above. Therefore, there are the
sets} $A$ \textit{and} $B\subseteq X,$ \textit{the H-function} $%
f:A\rightarrow X$ \textit{and a cut function} $\ f_{r}$ \textit{such that} 
\begin{equation*}
C=\{f_{r}:A\rightarrow \{0,1\}~/~\ f_{r}\left( x\right) =1,\text{\textit{if
\ and \ only \ if} \ }r\ast \ f\left( x\right) =\theta ,\forall x\in A,r\in
B\}.\Box 
\end{equation*}

\textbf{Proposition 3.9. }(Flaut, 2015, Remark 3.6.) \textit{If} $\mathfrak{N%
}_{n}$ \textit{is the number of all finite non-isomorphic BCK-algebras with} 
$\ n$ \textit{elements, then\ }$\mathfrak{N}_{n}\geq 2^{\frac{\left(
n-1\right) \left( n-2\right) }{2}}.\medskip $

\textbf{Proposition 3.10. } \textit{If }$\mathcal{N}$ \textit{is the number
of all finite non-isomorphic Hilbert algebras }$(X,\cdot ,\theta )$\textit{\
with} $n$ \textit{elements, then} $\mathcal{N}\geq 2^{\frac{\left(
n-1\right) \left( n-2\right) }{2}}.\smallskip $

\textbf{Proof.} From Remark 2.1, we know that the multiplication $"\ast "$
given in relation $\left( 2.1\right) \,$\ defines on the $(X,\ast ,\theta )$
a structure of a positive implicative BCK-algebra. Since to each positive
implicative BCK-algebra corresponds a Hilbert algebra, we will use
Proposition 2.4 and we obtain the required result.$\Box \medskip $

\textbf{Theorem 3.11.} \ \textit{Let} $V$ \textit{be a binary block-code
with }$n$\textit{\ codewords of length} $n$, $V=\{x_{1,}x_{2},...,x_{n}\}.$ 
\textit{We consider the codewords} $x_{1,}x_{2},...,x_{n}$ \textit{%
lexicographically ordered,} $x_{1}\geq _{lex}x_{2}\geq _{lex}...\geq
_{lex}x_{n}$ \textit{such that} $x_{1}=\underset{n}{\underbrace{11...1}}%
,x_{2}=\underset{n}{\underbrace{010...0}},x_{3}=\underset{n}{\underbrace{%
001...0}},x_{n}=\underset{n}{\underbrace{000...1}}.$ \textit{With the above
notations, the Hilbert algebra }$\left( X,\cdot ,\theta \right) ,$ \textit{%
obtained} \textit{as in Remark 3.2,} \textit{is a semisimple Hilbert
algebra.\smallskip }

\textbf{Proof.} The algebra $X$ ~has the following elements $X=\{(\theta
=x_{1}),x_{2},...,x_{n}\}.$ Using the construction of this algebra, we
obtain that $x_{i}\cdot x_{j}=x_{j},$ for all $i,j\in \{1,2,...,n\},i\neq j,$
therefore, for each $i\neq 1,$ we have that $L_{i}=\{x_{j}\in X~\shortmid
j\neq i\}$ is a maximal proper filter in $X.$ Since $\underset{i\neq 1}{\cap 
}L_{i}=\{\theta \},$ it results that $X$ is a semisimple Hilbert algebra.$%
\Box \medskip $

\textbf{Theorem 3.12. }\textit{Let} $V$ \textit{be a binary block-code with }%
$n$\textit{\ codewords of length} $n$, $V=\{x_{1,}x_{2},...,x_{n}\}.$ 
\textit{We consider the codewords} $x_{1,}x_{2},...,x_{n}$ \textit{%
lexicographically ordered,} $x_{1}\geq _{lex}x_{2}\geq _{lex}...\geq
_{lex}x_{n}.$ \textit{If the associated matrix }$M_{V}=\left( m_{i,j}\right)
_{i,j\in \{1,2,...,n\}}\in \mathcal{M}_{n}(\{0,1\})~$\textit{of} \textit{the
code} $V$ \textit{is upper triangular, with} $m_{ii}=1,$ \textit{for all} $%
i\in \{1,2,...,n\},$ \textit{and the last column is equal} \textit{to} $%
\underset{n}{\underbrace{11...1}}$ , \textit{with the above notations, the
Hilbert algebra }$\left( X,\cdot ,\theta \right) ,$ \textit{obtained} 
\textit{as in Remark 3.2,} \textit{\ is a local Hilbert algebra.\smallskip }

\textbf{Proof.} The algebra $X$ has the following elements $X=\{(\theta
=x_{1}),x_{2},...,x_{n}\}.$ Using the construction of this algebra, since $%
x_{i}\cdot x_{j}=x_{j}$ or $\theta ,$ and $x_{n}\cdot x_{j}=\theta ,$ for
all $j\in \{1,2,...,n\},$ we obtain that $L=\{\theta ,x_{2},...,x_{n-1}\}$
is the only maximal proper filter in $X,$ therefore $X$ is a local Hilbert
algebra.$~\Box \medskip $ 
\begin{equation*}
\end{equation*}

\textbf{4}. \textbf{Examples}%
\begin{equation*}
\end{equation*}

\textbf{Example 4.1. }\ i) Let $C=\{0000,0001,0010,0011\}=%
\{w_{6},w_{7},w_{8},w_{9}\}$ be a linear binary block code and let $%
X=\{\theta ,w_{2},w_{3},w_{4},w_{5},w_{6},w_{7},w_{8},w_{9}\}$ be the
obtained BCK-algebra. The multiplication "$\ast $" of this algebra is given
in the table below ( see (Borumand Saeid, Fatemidokht, Flaut, Kuchaki
Rafsanjani, 2015)):\medskip 

\begin{tabular}{|l|l|l|l|l|l|l|l|l|l|}
\hline
$\ast $ & $\theta $ & $w_{2}$ & $w_{3}$ & $w_{4}$ & $w_{5}$ & $w_{6}$ & $%
w_{7}$ & $w_{8}$ & $w_{9}$ \\ \hline
$\theta $ & $\theta $ & $\theta $ & $\theta $ & $\theta $ & $\theta $ & $%
\theta $ & $\theta $ & $\theta $ & $\theta $ \\ \hline
$w_{2}$ & $w_{2}$ & $\theta $ & $w_{2}$ & $w_{2}$ & $w_{2}$ & $\mathbf{w}_{2}
$ & $\mathbf{w}_{2}$ & $\mathbf{\theta }$ & $\mathbf{\theta }$ \\ \hline
$w_{3}$ & $w_{3}$ & $w_{3}$ & $\theta $ & $w_{3}$ & $w_{3}$ & $\mathbf{w}_{3}
$ & $\mathbf{w}_{3}$ & $\mathbf{\theta }$ & $\mathbf{w}_{3}$ \\ \hline
$w_{4}$ & $w_{4}$ & $w_{4}$ & $w_{4}$ & $\theta $ & $w_{4}$ & $\mathbf{w}_{4}
$ & $\mathbf{w}_{4}$ & $\mathbf{w}_{4}$ & $\mathbf{\theta }$ \\ \hline
$w_{5}$ & $w_{5}$ & $w_{5}$ & $w_{5}$ & $w_{5}$ & $\theta $ & $\mathbf{w}_{5}
$ & $\mathbf{w}_{5}$ & $\mathbf{w}_{5}$ & $\mathbf{w}_{5}$ \\ \hline
$w_{6}$ & $w_{6}$ & $w_{6}$ & $w_{6}$ & $w_{6}$ & $w_{6}$ & $\theta $ & $%
w_{6}$ & $w_{6}$ & $w_{6}$ \\ \hline
$w_{7}$ & $w_{7}$ & $w_{7}$ & $w_{7}$ & $w_{7}$ & $w_{7}$ & $w_{7}$ & $%
\theta $ & $w_{7}$ & $w_{7}$ \\ \hline
$w_{8}$ & $w_{8}$ & $w_{8}$ & $w_{8}$ & $w_{8}$ & $w_{8}$ & $w_{8}$ & $w_{8}$
& $\theta $ & $w_{8}$ \\ \hline
$w_{9}$ & $w_{9}$ & $w_{9}$ & $w_{9}$ & $w_{9}$ & $w_{9}$ & $w_{9}$ & $w_{9}$
& $w_{9}$ & $\theta $ \\ \hline
\end{tabular}%
\medskip 

The multiplication table for the obtained Hilbert algebra is\medskip

\begin{tabular}{|l|l|l|l|l|l|l|l|l|l|}
\hline
$\cdot $ & $\theta $ & $w_{2}$ & $w_{3}$ & $w_{4}$ & $w_{5}$ & $w_{6}$ & $%
w_{7}$ & $w_{8}$ & $w_{9}$ \\ \hline
$\theta $ & $\theta $ & $w_{2}$ & $w_{3}$ & $w_{4}$ & $w_{5}$ & $w_{6}$ & $%
w_{7}$ & $w_{8}$ & $w_{9}$ \\ \hline
$w_{2}$ & $\theta $ & $\theta $ & $w_{3}$ & $w_{4}$ & $w_{5}$ & $w_{6}$ & $%
w_{7}$ & $w_{8}$ & $w_{9}$ \\ \hline
$w_{3}$ & $\theta $ & $w_{2}$ & $\theta $ & $w_{4}$ & $w_{5}$ & $w_{6}$ & $%
w_{7}$ & $w_{8}$ & $w_{9}$ \\ \hline
$w_{4}$ & $\theta $ & $w_{2}$ & $w_{3}$ & $\theta $ & $w_{5}$ & $w_{6}$ & $%
w_{7}$ & $w_{8}$ & $w_{9}$ \\ \hline
$w_{5}$ & $\theta $ & $w_{2}$ & $w_{3}$ & $w_{4}$ & $\theta $ & $w_{6}$ & $%
w_{7}$ & $w_{8}$ & $w_{9}$ \\ \hline
$w_{6}$ & $\theta $ & $\mathbf{w}_{2}$ & $\mathbf{w}_{3}$ & $\mathbf{w}_{4}$
& $\mathbf{w}_{5}$ & $\theta $ & $w_{7}$ & $w_{8}$ & $w_{9}$ \\ \hline
$w_{7}$ & $\theta $ & $\mathbf{w}_{2}$ & $\mathbf{w}_{3}$ & $\mathbf{w}_{4}$
& $\mathbf{w}_{5}$ & $w_{6}$ & $\theta $ & $w_{8}$ & $w_{9}$ \\ \hline
$w_{8}$ & $\theta $ & $\mathbf{\theta }$ & $\mathbf{\theta }$ & $\mathbf{w}%
_{4}$ & $\mathbf{w}_{5}$ & $w_{6}$ & $w_{7}$ & $\theta $ & $w_{9}$ \\ \hline
$w_{9}$ & $\theta $ & $\mathbf{\theta }$ & $\mathbf{w}_{3}$ & $\mathbf{%
\theta }$ & $\mathbf{w}_{5}$ & $w_{6}$ & $w_{7}$ & $w_{8}$ & $\theta $ \\ 
\hline
\end{tabular}%
\medskip 

Using above notations, for $B=\{w_{6},w_{7},w_{8},w_{9}\}$ and $%
A=\{w_{2},w_{3},w_{4},w_{5}\},$ we remark that we obtain the initial code.
We \ remark that \newline
$L_{1}=\{\theta ,w_{2},w_{3},w_{4},w_{5},w_{6},w_{7},w_{9}\},L_{2}=\{\theta
,w_{2},w_{3},w_{4},w_{5},w_{6},w_{7},w_{8}\},$\newline
$L_{3}=\{\theta ,w_{2},w_{3},w_{4},w_{5},w_{6},w_{8},w_{9}\},L_{4}=\{\theta
,w_{2},w_{3},w_{4},w_{5},w_{7},w_{8},w_{9}\},$\newline
$L_{5}=\{\theta ,w_{2},w_{3},w_{4},w_{6},w_{7},w_{8},w_{9}\},L_{6}=\{\theta
,w_{2},w_{3},w_{5},w_{6},w_{7},w_{8},w_{9}\}$ are all maximal filters. Since 
$\underset{i=1}{\overset{6}{\cap }}L_{i}=\{\theta ,w_{2},w_{3}\},$ it
results that this algebra is not semisimple and it is not local.\medskip 

\textbf{Example 4.2.} \ We consider the binary block code\newline
$C=\{11111,01011,00111,00011,00001\}.$\newline
Since the codewords are lexicographically ordered, the obtained BCK algebra
is $H=\{\theta ,a,b,c,d\},\left( H,\ast \right) ,$ the obtained Hilbert
algebra is $\left( H,\cdot \right) $ and have the multiplication given in
the tables below.\medskip 

\begin{tabular}{|l|l|l|l|l|l|}
\hline
$\ast $ & $\theta $ & $a$ & $b$ & $c$ & $d$ \\ \hline
$\theta $ & $\theta $ & $\theta $ & $\theta $ & $\theta $ & $\theta $ \\ 
\hline
$a$ & $a$ & $\theta $ & $a$ & $\theta $ & $\theta $ \\ \hline
$b$ & $b$ & $b$ & $\theta $ & $\theta $ & $\theta $ \\ \hline
$c$ & $c$ & $c$ & $c$ & $\theta $ & $\theta $ \\ \hline
$d$ & $d$ & $d$ & $d$ & $d$ & $\theta $ \\ \hline
\end{tabular}%
\ \ \ \ \ \ \ \ \ 
\begin{tabular}{|l|l|l|l|l|l|}
\hline
$\cdot $ & $\theta $ & $a$ & $b$ & $c$ & $d$ \\ \hline
$\theta $ & $\theta $ & $a$ & $b$ & $c$ & $d$ \\ \hline
$a$ & $\theta $ & $\theta $ & $b$ & $c$ & $d$ \\ \hline
$b$ & $\theta $ & $a$ & $\theta $ & $c$ & $d$ \\ \hline
$c$ & $\theta $ & $\theta $ & $\theta $ & $\theta $ & $d$ \\ \hline
$d$ & $\theta $ & $\theta $ & $\theta $ & $\theta $ & $\theta $ \\ \hline
\end{tabular}%
\medskip 

The proper filters are $L_{1}=\{\theta ,a\},L_{2}=\{\theta
,b\},L_{3}=\{\theta ,a,b\},L_{4}=\{\theta ,a,b,c\}.$\newline
$L_{4}$ is the only maximal filter, therefore $\left( H,\cdot \right) $ is a
local Hilbert algebra (see Theorem 3.12). $H$ is not a semisimple
algebra.\medskip

\textbf{Example 4.3.} We consider the binary block code $C=%
\{1111,0100,0010,0001\}.$\newline
Since the codewords are lexicographically ordered, the obtained BCK algebra
is $H=\{\theta ,a,b,c\},\left( H,\ast \right) ,$ the obtained Hilbert
algebra is $\left( H,\cdot \right) $ and have the multiplication given in
the tables below.\medskip 

\begin{tabular}{|l|l|l|l|l|}
\hline
$\ast $ & $\theta $ & $a$ & $b$ & $c$ \\ \hline
$\theta $ & $\theta $ & $\theta $ & $\theta $ & $\theta $ \\ 
$a$ & $a$ & $\theta $ & $a$ & $a$ \\ \hline
$b$ & $b$ & $b$ & $\theta $ & $b$ \\ \hline
$c$ & $c$ & $c$ & $c$ & $\theta $ \\ \hline
\end{tabular}
\ \ \ \ \ 
\begin{tabular}{|l|l|l|l|l|}
\hline
$\cdot $ & $\theta $ & $a$ & $b$ & $c$ \\ \hline
$\theta $ & $\theta $ & $a$ & $b$ & $c$ \\ \hline
$a$ & $\theta $ & $\theta $ & $b$ & $c$ \\ \hline
$b$ & $\theta $ & $a$ & $\theta $ & $c$ \\ \hline
$c$ & $\theta $ & $a$ & $b$ & $\theta $ \\ \hline
\end{tabular}%
\medskip

The proper maximal filters are $L_{1}=\{\theta ,a,b\},L_{2}=\{\theta
,b,c\},L_{3}=\{\theta ,a,c\}.L_{1},$ $L_{2},L_{3}$ are maximal filters and $%
L_{1}\cap L_{2}\cap L_{3}=\{e\},$ therefore $\left( H,\cdot \right) $ is not
a local Hilbert algebra but it is a semisimple algebra (see Theorem 3.11).

\begin{equation*}
\end{equation*}

\textbf{Conclusions. }In the papers (Jun, Song, 2011), (Flaut, 2015),
(Borumand Saeid, Fatemidokht, Flaut, Kuchaki Rafsanjani, 2015), some
connections between BCK-algebras and binary block codes were described. In
this paper we make some connections between Hilbert algebras and binary
block codes via BCK-algebra defined by relation (2.1). In this way, using
codes, we found examples of classes of semisimple Hilbert algebras and
classes of local Hilbert algebras. As further research, we will try to find
answers to the reverse problem, namely, how properties of BCK-algebras,
BCI-algebras, Hilbert algebras can influence the properties of binary block
codes.

\begin{equation*}
\end{equation*}

\textbf{References}%
\begin{equation*}
\end{equation*}

Borumand Saeid, A., Fatemidokht, H., Flaut, C., Kuchaki Rafsanjani, M.,
(2015), \textit{On Codes based on BCK-algebras,} arxiv, 13 pages.

D. Bu\c{s}neag, \textit{On the maximal deductive system of a bounded Hilbert
algebra}, (1987), Bull. Math. Soc. Sci. Math. Roumanie, 31(1)(79), pp. 9-21.

Chajda, I., Hala\v{s}, R.,  Jun, Y.B.,  (2002), \textit{Annihilators and
deductive systems in commutative Hilbert algebras}, Comment. Math. Univ.
Carolin., 43(3), pp. 407--417.

Dan, C., (2008), \textit{Hilbert algebras of Fractions},  Int. J. Math. and
Math. Sci., 2009, Article ID 589830 (16 pages).

Diego A., (1966), \textit{Sur les alg\'{e}bras de Hilbert}, Ed. Hermann, Coll%
\'{e}ction de Logique Math. Serie A, 21, pp. 1--52.

Dudek, W.A.,\ (1999), \textit{On embedding Hilbert algebras in BCK-algebras}%
, Mathematica Moravica, 3, pp. 25-28.

Flaut, C., (2015), \textit{BCK-algebras arising from block codes}, J.
Intell. Fuzzy Syst., 28(4), pp. 1829-1833.

Jun, Y. B.,  Song, S. Z.,  (2011), \textit{Codes based on BCK-algebras},
Inform. Sciences., 181, pp.  5102-5109.

Piciu, D., Bu\c{s}neag, C., (2010), \textit{The localization of commutative
(unbounded) Hilbert algebras}, Math. Rep., 3(12)(62), 16 pages\textbf{.} 
\begin{equation*}
\end{equation*}

\end{document}